\documentclass[aps,twocolumn,pre]{revtex4}
\usepackage{epsfig}
\usepackage{dcolumn}
\usepackage{times}
\usepackage{mathptm}

\begin{document}

% common abbreviations
\newcommand{\beq}[0]{\begin{equation}}
\newcommand{\eeq}[0]{\end{equation}}
\newcommand{\di}[0]{d}
\newcommand{\mulame}[0]{\mu_L}
\newcommand{\lamlame}[0]{\lambda_L}
\newcommand{\ups}[0]{\upsilon}
\newcommand{\half}[0]{{\textstyle \frac{1}{2}}}
\newcommand{\inv}[1]{{\textstyle \frac{1}{#1}}}
\newcommand{\breuk}[2]{{\textstyle \frac{#1}{#2}}}
\newcommand{\deriv}[2]{\frac{\partial #1}{\partial #2}}
\newcommand{\derivtwo}[2]{\frac{\partial^2 #1}{\partial {#2}^2}}

% for this article
\newcommand{\yrem}[0]{y_{\mathrm{rem}}}
\newcommand{\yadd}[0]{y_{\mathrm{add}}}
\newcommand{\yshrink}[0]{y_{\mathrm{shrink}}}
\newcommand{\ygrow}[0]{y_{\mathrm{grow}}}
\newcommand{\logpins}[0]{-k_BT\ln P_\mathrm{ins}}

\newcommand{\sigmared}[0]{\frac{\sigma}{\langle \sigma \rangle}}
\newcommand{\sigmaredl}[0]{\sigma/\langle \sigma \rangle}
\newcommand{\msigma}[0]{\langle\sigma\rangle}

\title{Large effect of polydispersity on defect concentrations in
  colloidal crystals}

\author{Sander Pronk} 
\author{Daan Frenkel}
%\email{frenkel@amolf.nl} 
\affiliation{
  FOM Institute for Atomic and Molecular Physics\\
  Kruislaan 407\\
  1098 SJ Amsterdam\\
  the Netherlands}

\begin{abstract}
  We compute the equilibrium concentration of stacking faults and
  point defects in polydisperse hard-sphere crystals. We find that,
  while the concentration of stacking faults remains similar to that
  of monodisperse hard sphere crystals, the concentration of vacancies
  decreases by about a factor two.  Most strikingly, the concentration
  of interstitials in the maximally polydisperse crystal may be some
  six orders of magnitude larger than in a monodisperse crystal. We
  show that this dramatic increase in interstitial concentration is
  due to the increased probability of finding small particles and that
  the small-particle tail of the particle size distribution is crucial
  for the interstitial concentration in a colloidal crystal.
\end{abstract}

\maketitle

\section{Introduction}
The experimental study of colloidal crystals is of interest for at
least two reasons. First of all, the possibility to design the
constituents of such crystals, allows us to gain insight into the
factors that determine the structure and kinetics of formation of
crystalline materials. In addition, colloidal crystals are of interest
because of their potential application as photonic
materials\cite{vanBlaaderenPhotonic}.  To a first approximation, one
might view colloidal crystals as scale models of atomic crystals. But
this analogy is flawed for several reasons. First of all, the
intermolecular forces between colloidal particles may be qualitatively
different from those between atoms.  Secondly, the dynamics of
colloidal matter is intrinsically different from that of atomic
materials, due to the presence of a solvent.  Finally, unlike atomic
materials, colloidal systems are never completely monodisperse. This
polydispersity may have important consequence for the phase behavior
and structural properties of the colloidal crystals. In addition,
polydispersity can have an effect on the equilibrium concentration of
(point) defects in colloidal crystals. As defects may strongly
influence the photonic properties of colloidal crystals, a better
understanding of the effect of polydispersity on defect
concentrations, may also be of practical relevance for the design of
photonic crystals.

In the present paper, we describe a numerical study of the effect of
polydispersity on the concentration of stacking faults, vacancies and
interstitials in hard-sphere colloidal crystals.

\section{Simulation Methods}
\subsection{Semigrand Canonical Ensemble}

To simulate a the equilibrium properties of polydisperse hard-sphere
crystals, we used the semigrand canonical ensemble
method~\cite{KofkeSemigrand,BolhuisPdisp1}. For a system with
continuous size polydispersity, the free-energy functional of the
semigrand canonical ensemble is given by:
\begin{eqnarray}
 Y(N,P,T,\sigma_0,\{\Delta\mu\}) = U - TS + PV + N\mu(\sigma_0) 
& &
  \nonumber \\
  - N \int \di \sigma \left[ \mu(\sigma) - \mu(\sigma_0) \right]
  p(\sigma)  & &
\end{eqnarray}
where $N$ is the total number of particles in the system, $P$ is
the pressure, $T$ is the temperature and the set $\{\Delta\mu\}$
denotes the differences between $\mu(\sigma)$, the chemical
potential of a species with diameter $\sigma$, and
$\mu(\sigma_0)$, the chemical potential of an (otherwise
arbitrary) reference species: $\Delta\mu(\sigma)\equiv\mu(\sigma)
- \mu(\sigma_0)$. As we are dealing with hard-core particles, we
choose our unit of energy to be equal to $k_BT$. $p(\sigma)$
denotes the probability of finding a particle with diameter
$\sigma$. The set of thermodynamic fields $\{\Delta\mu\}$ act as
control parameters that determine the particle-size distribution.
In the present work, we assume a quadratic dependence of
$\Delta\mu(\sigma)$ on $\sigma-\sigma_0$:
\begin{equation}
\beta\left[\mu(\sigma) - \mu(\sigma_0)\right]  = - ( \sigma - \sigma_0)^2/2\nu
\label{pdd:mufunc}
\end{equation}
Where $\beta \equiv 1/k_BT$. The parameter $\nu$ determines the degree
of polydispersity. At infinite dilution, the size distribution is
directly given by $p(\sigma)=c\exp(- ( \sigma - \sigma_0)^2/2\nu)$. At
finite concentrations, the size distribution cannot be inferred
directly from the functional form of $\Delta\mu(\sigma)$. Both the
average particle diameter and the actual polydispersity $s$ (defined
through $s^2 \equiv \langle \sigma^2 \rangle/\langle \sigma \rangle^2
- 1$) must be determined in the semigrand ensemble simulations. Once
the functional form of $\Delta\mu(\sigma)$ has been specified, the
semi-grand partition function $\Xi$ is a function of $N,P,T,\nu$ and
$\sigma_0$.
\begin{equation}
  \begin{array}{rr}
\lefteqn{  \Xi(N,P,T,\nu,\sigma_0)  = } & 
  \nonumber \\
& \int \di V \int \di \mathbf{r}^{N} \int \di\sigma^{N} \exp \bigg(
      -\beta \left[ PV + U \left(r^{N}, \sigma^{N} \right) \right]  
   \nonumber \\
&    {}  -  \sum_{i} \frac{ (\sigma_i - \sigma_0)^2 }{2\nu}
    \bigg) 
    \label{pdd:partsum1}
  \end{array}
\end{equation}
The semigrand free energy $Y$ is related to $\Xi$ through
$Y=-k_BT\ln\Xi$.  To sample the configurations of the semi-grand
ensemble, we use Metropolis-style Monte Carlo sampling of all
variables that characterize a given configuration of the $N$-particle
system.  In addition to the usual trial moves that attempt to change
the particle coordinates $\{\mathbf{r}^N\}$ and the system volume $V$,
there are trial moves to change the diameter of a particle. As has
been explained by Bolhuis and Kofke, it is computationally more
efficient to combine volume-changing moves with particle resizing
moves~\cite{BolhuisPdisp1}.

To calculate the chemical potential of the reference species,
thermodynamic integration was used. As a reference state, we took the
monodisperse hard-sphere crystal near coexistence, for which the free
energy per particle is accurately known~\cite{PolsonEinstein}. In
order to compute the change in free energy with $P$ and $\nu$, we make
use of the following thermodynamic relations:
\begin{eqnarray}
  \left( \frac{\partial Y}{\partial P} \right)_{N,T,\sigma_0,\nu} & = & V
  \nonumber \\
  \left( \frac{\partial Y}{\partial \nu} \right)_{N,P,T,\sigma_0} & = &
  N \int \di \sigma ' p(\sigma ') \frac{-(\sigma ' - \sigma_0)^2}{2\nu^2}
  \label{pdd:tidiff}
\end{eqnarray}
The semigrand free energy of an ideal, non-interacting system of
polydisperse particles, is
\begin{eqnarray}
  Y_{\mathrm{id}} & = & -k_BT \ln \int \di V \exp(-\beta PV)
\nonumber \\ & & 
  \int \di \mathbf{r}^N
 \int \di
  \sigma^N \exp
  \left( -  \sum_{i}
    \frac{ (\sigma_i - \sigma_0)^2 }{2\nu} \right) \nonumber \\
   & = &
 N\mu_{\mathrm{id}} (\sigma_0) =
  G_{\mathrm{id}}- \frac{Nk_BT}{2} \ln (2 \pi \nu)
\end{eqnarray}
We can now employ the following scheme to compute
$\mu_{\mathrm{ex}}(\sigma_0)$ by thermodynamic integration, using as
input our knowledge of the excess chemical potential
$\mu_{\mathrm{ex},0}$ of a monodisperse hard sphere system at pressure
$P_0$:
\begin{eqnarray}
  \mu_{\mathrm{ex}}(\sigma_0) &  = & \mu_{\mathrm{ex},0} +
   \frac{1}{N} \int_{P_0}^{P} \di P'
   \left\langle
     V  - \frac{(N+1)k_BT}{P'}
  \right\rangle
  \nonumber \\
  & & + \frac{1}{N} \int_{0}^{\nu}
  \di \nu ' \left\langle \frac{-\sum_i (\sigma_i - \sigma_0)^2}{2\nu '^2} 
    + \frac{Nk_BT}{2\nu '}
  \right \rangle \; .
\end{eqnarray}

\subsection{Interstitial Concentration}

The methods that we used to calculate the concentration of point
defects are similar to those discussed in
Ref.~\onlinecite{PronkFrenkelPointDefects}.  We first consider the
free energy $Y_{M,n_V,n_I}$ of a crystalline system containing $M$
lattice sites, $n_V$ vacancies and $n_I$ interstitials. The total
number of particles in this system is $N=M+n_I-n_V$. It is convenient
to consider interstitials and vacancies separately.
 
By analogy to the derivation of interstitial concentrations in
monodisperse systems\cite{PronkFrenkelPointDefects}, it is
straightforward to show that the concentration of interstitials
($x_I$) is given by $x_I \approx \exp(-\beta y_I)$, where $y_I$ is
defined as $y_I=Y_{M,0,1}-Y_{M+1,0,0}$. It is convenient to rewrite
$y_I$ as
\begin{eqnarray}
  y_I & = & Y_{M,0,1} - Y_{M+1,0,0}
  \nonumber \\
  & = & Y_{M,0,1} - Y_{M,0,0} + Y_{M,0,0} - Y_{M+1,0,0}
  \nonumber \\
  & = & Y_{M,0,1} - Y_{M,0,0} -  
  \Big[ \mu_{\mathrm{id}}(\sigma_0) + \mu_{\mathrm{ex}}(\sigma_0)\Big]
  \nonumber \\
  & = & Y_{M,0,1} - \Big[ Y_{M,0,0} +  \mu_{\mathrm{id}}(\sigma_0) \Big]
  - \mu_{\mathrm{ex}}(\sigma_0)
  \nonumber \\
  & = & \yadd - \mu_{\mathrm{ex}}(\sigma_0)
  \label{pdd:yint}
\end{eqnarray}
Here $\yadd$ is the free energy difference between a system with one
interstitial and a perfect crystal plus one ideal (non-interacting)
particle. The quantity $Y_{M+1,0,0}$, the free energy of a system with
$M+1$ lattice sites and no defects, is an abstract quantity that does
not neccesarily correspond to a crystal with realizable lattice in an
orthorhombic simulation box; its value is well-defined by virtue of
the extensivity of free energy.

To calculate $\yadd$, we simulate a crystal with $M$ lattice sites and
$M+1$ particles, of which particle $j$ has a scaled hard-core diameter
$a\sigma_j$.  The diameter scaling parameter $a$ can be varied during
the simulation, so that we sample the partition defined by
\begin{eqnarray}
\lefteqn{\Xi'_{M,0,1}(M+1,P,T,\sigma_0,\nu)  = } & &
\nonumber \\
& & \int_0^1 \di a \ \Xi_{M,0,1}(M+1,P,T,\sigma_0,\nu,a) 
\end{eqnarray}
where $\Xi_{M,0,1}(M+1,P,T,\nu,\sigma_0,a)$ is defined as in
Eq.~\ref{pdd:partsum1}, but with configurational energy
$U(\mathbf{r}^M, \sigma^{M}, a \sigma_j)$. We stress that particle $j$
differs from the other particles only in the overlap criterion, not in
the probability distribution that determines diameter sampling: for
the overlap criterion, the particle radius of this particle is
$a\sigma$, whereas its weight in the Semigrand chemical potential
distribution of Eq.~\ref{pdd:mufunc} is still determined by $\sigma$.

During the simulation, we construct a histogram $P(a|M+1,P,T,\nu)$:
\begin{eqnarray}
\lefteqn{P(a|M+1,P,T,\nu) = } & & \nonumber \\
& &  \frac{\int_0^1 \di a' \delta(a-a') 
  \Xi_{M,0,1}(M+1,P,T,\nu,\sigma_0,a)}
{\Xi'_{M,0,1}(M+1,P,T,\nu,\sigma_0)}
\end{eqnarray}
With this histogram we can calculate
\begin{equation}
  \ygrow =  -k_BT \ln \frac{P(a=1|M+1,P,T,\nu,\sigma_0)}{P(a=0|M+1,P,T,\nu,
    \sigma_0)}
  \label{pdd:grow}
\end{equation}
where $\ygrow$ is the reversible work needed to transform an
interacting point particle ($a$=0) into a particle with a hard-core
diameter $\sigma_j$ (corresponding to $a$=1). In order to sample the
full range of $a$-values from 0 to 1, it is necessary to use biased
sampling. We employed multicanonical/umbrella sampling
\cite{BergNeuhausMulticanonical,SmithBruceMulticanonical} to generate
$P(a|M,P,T,\nu,\sigma_0)$.
 
To obtain the total interstitial free energy $\yadd$ we must still add
the free energy change associated with the transformation of a
non-interacting particle into an interacting point particle. This free
energy change is determined by the ratio of the volumes accessible to
the two types of particles:
\begin{eqnarray}
  \yadd - \ygrow & = & -k_BT \ln \frac{\langle V_{\mathrm{acc}}\rangle}{V}
  \nonumber \\
  & = & - k_BT \ln \langle 1-\eta \rangle
\end{eqnarray}
where $V_{\mathrm{acc}}$ is the volume accessible to the point
particle and $\eta$ denotes the volume fraction of the defect-free
hard-sphere crystal.  It is not necessary to confine the interstitial
to a particular Wigner-Seitz cell, as interstitials diffuse quickly
through the system. If this were not the case, both the scaled and the
unscaled particle would have to be confined to a particular
Wigner-Seitz cell (or even, to one particular interstitial cavity).

\subsection{Vacancy Concentration}

For the vacancies, we can get for the concentration $x_V \approx \exp
(-\beta y_V)$ (see Ref.~\onlinecite{PronkFrenkelPointDefects}), with
$y_V=Y_{M+1,1,0}-Y_{M,0,0}$ and introduce the analogous free energy to
$\yadd$:
\begin{eqnarray}
y_V & = & Y_{M+1,1,0}-Y_{M,0,0}
\nonumber \\
    & = & Y_{M+1,1,0} - Y_{M+1,0,0} + Y_{M+1,0,0} - Y_{M,0,0}
\nonumber \\
    & = & Y_{M+1,1,0} - Y_{M+1,0,0} + \mu(\sigma_0)
\nonumber \\
    & = & Y_{M+1,1,0}  - Y_{M+1,0,0} +
    \mu_{\mathrm{id}}(\sigma_0)  + \mu_{\mathrm{ex}}(\sigma_0)
\nonumber \\
    & = & - \Big(  Y_{M+1,0,0} - \Big[ Y_{M+1,1,0}  +
    \mu_{\mathrm{id}}\left(\sigma_0\right) \Big] \Big) 
     + \mu_{\mathrm{ex}}(\sigma_0)
\nonumber \\
    & = & -\yrem + \mu_{\mathrm{ex}}(\sigma_0)
\end{eqnarray}
In this case, $\yrem$ is the free energy difference between a perfect
crystal and a crystal with one vacancy plus a non-interacting
particle.

If we assume that we can sample a system which can switch one particle
between being a normal particle ($b=b_n$) and a non-interacting
particle ($b=b_i$), we can introduce the equilibrium probability
$P(b|M,P,T,\nu,\sigma_0)$:
\begin{eqnarray}
\yrem & = & -k_BT \ln
\frac{P(b_n|M,P,T,\nu,\sigma_0)}{P(b_i|M,P,T,\nu,\sigma_0)} 
\nonumber \\
& = &
-k_BT \ln \frac{ \langle \pi(b_i \to b_n) \rangle}
{\langle \pi(b_n \to b_i)\rangle}
\label{pdd:switch}
\end{eqnarray}
where $\langle \pi(b_i \to b_n) \rangle$ is the mean transition
probability from $b=b_i$ to $b=b_n$. Because a real particle can
always switch to a non-interacting, particle, we can reduce the
expression for $\yrem$ to
\begin{equation}
\yrem = -k_BT \ln \langle \pi(b_i \to b_n) \rangle
\end{equation}
Now $\langle \pi(b_i \to b_n) \rangle$, the transition probability
from a state of a system with a vacancy and a non-interacting particle
to a perfect crystal, is related to the probability $P_\mathrm{ins}$
for the insertion of a (normal polydisperse) particle into the
%Wigner-Seitz cell associated with a 
vacancy:
\begin{equation}
-k_BT \ln \langle \pi(b_i \to b_n) \rangle= -k_BT ( \ln P_\mathrm{ins} )
\label{pdd:pins}
\end{equation}

In practice, the simulation will consist of a collection of $M-1$
normal particles and one ideal polydisperse particle which we keep in
the Wigner-Seitz cell of the tracked vacancy. We then do
multicanonical sampling, biasing on the number of overlaps that the
ideal particle would create if it would be switched to a real
particle, and get $P_\mathrm{ins}$ from the probability to create zero
overlaps. This scheme is essentially identical to that of Bennett and
Alder\cite{BennettAlderVacancies}, save for the multicanonical
sampling.

\section{Results}

\begin{figure}
\center
%\vspace{0.2cm}
  \includegraphics[width=0.95\linewidth]{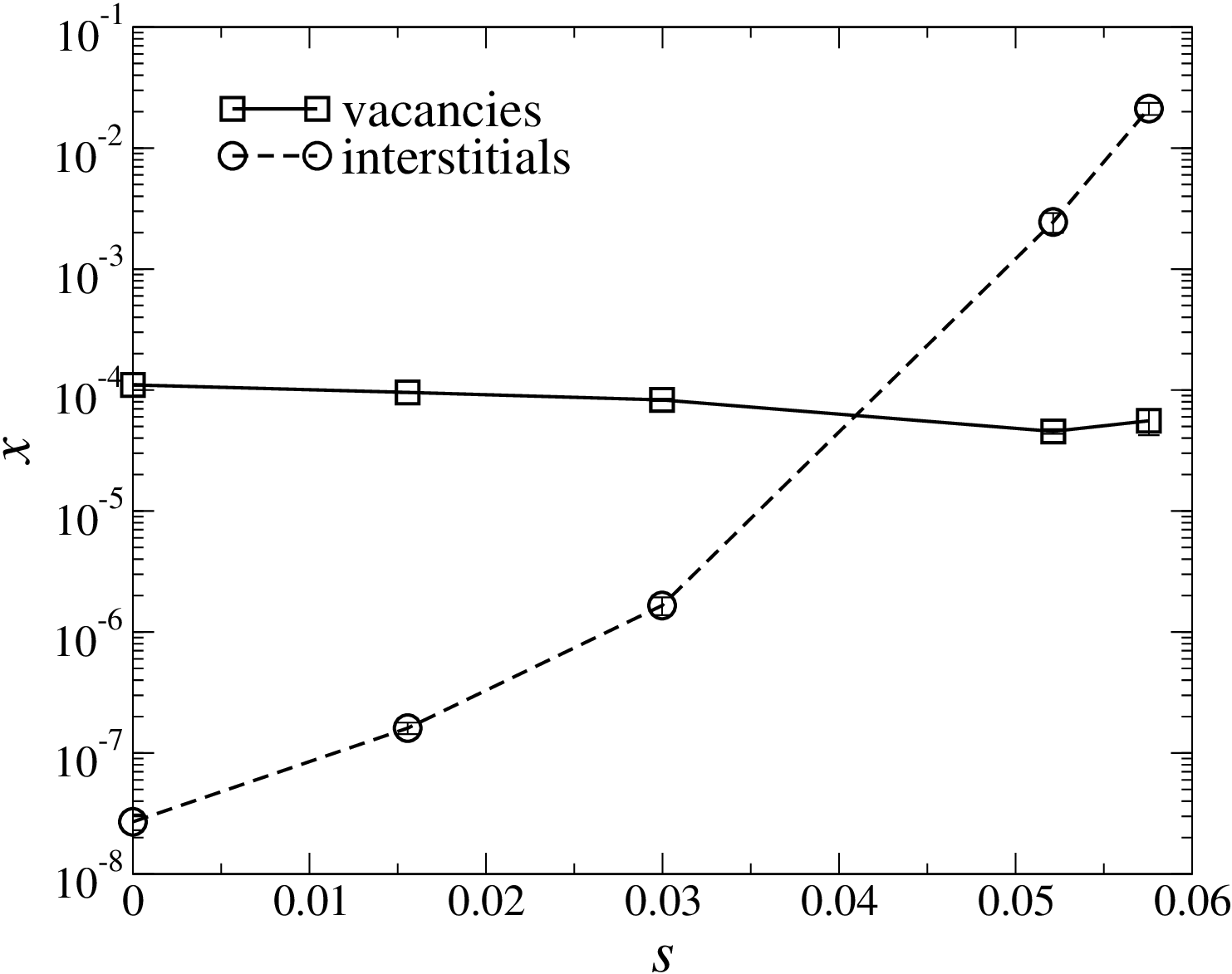}
  \caption{Point defect concentration ($x$) versus polydispersity
    ($s$)%, defined in Eq.~\ref{pdd:pdispdef}
    }
  \label{pdd:concgraph}
\end{figure}

The simulations to calculate the point defect concentration were done
at various points along the melting line of polydisperse hard sphere
crystals, as taken from Ref.~\onlinecite{BolhuisPdisp1}. The points
chosen give a polydispersity of approx. $1.5\%$, $3\%$, $5\%$ and
$5.8\%$.  The latter value corresponds to the maximum polydispersity
attainable with the chemical potential difference function used. Here,
the polydispersity $s$ is defined as the normalized second moment of
the particle diameter distribution
\begin{equation}
  s \equiv \frac{\sqrt{\langle \sigma^2 \rangle - \langle \sigma \rangle^2}}
  {\langle \sigma \rangle}
  \label{pdd:pdispdef}
\end{equation}

All simulations were performed on $256 (\pm 1)$ particle systems (a
cubic fcc $4\times4\times4$ lattice); a simulation of a larger system
in the monodisperse case in Ref.~\onlinecite{PronkFrenkelPointDefects}
shows that this particle number is sufficient for the required
accuracy. For the (interstitial) calculation of $\ygrow$, the
$P(a|M+1,P,T)$ histograms were divided into $5$ windows for which
simulations were run in parallel. The multicanonical biasing weights
were generated starting with the weights for the monodisperse case and
took $10$ -- $80$ runs of $4 \cdot 10^5$ MC sweeps (Monte Carlo cycles
per particle) per CPU to converge.  The final results were obtained
using typically $80$ runs of $4 \cdot 10^5$ sweeps per CPU.  In the
case of vacancies there was one window for which about $20$ runs of $1
\cdot 10^6$ sweeps were needed to equilibrate the weights after which
about $40$ runs of similar length were done for the final results.
The equilibrium concentration of the two types of point vacancies as a
function of different polydispersities is shown in Table
\ref{pdd:restab} and Fig.~\ref{pdd:concgraph}.

The values of $\mu_{\mathrm{ex}}(\sigma_0)$, required for both the
vacancy and interstitial concentration, were calculated using
thermodynamic integration using the free energy differentials of
Eq.~\ref{pdd:tidiff}. Integration was done along the $P$-$\nu$ points
shown in Table \ref{pdd:restab}, with $20$ steps between each step and
$1 \cdot 10^6$ averaging sweeps per step.

\begin{table*}[tb]
  \begin{center}
    \begin{tabular}{|l|l|l|l|l|l|l|l|l|l|l|l|}
      \hline
      \hline
      $\nu$ & 0 & 0.00025 & 0.001 & 0.004 & 0.0056 \\
      $P$ & 11.7 & 12.08 & 13.56 & 26.9 & 82.6 \\
      \hline
%$1 - \langle \sigma \rangle^2/\langle \sigma^2 \rangle$ 
      $\eta$ & 0.54329 & 0.54522(8) & 0.54641(6) & 0.55726(6) & 0.56997(6) \\
      $\langle \sigma \rangle$& 1 & 0.992 & 0.967 & 0.815 & 0.589 \\
      $s$ & 0 & 0.015562(3) & 0.029974(7) &  0.05213(3) & 0.05755(5) \\
      $\mu_{\mathrm{ex}}$ & 17.071 & 17.418 & 18.308 & 24.350 & 37.516 \\
      $\mu_{\mathrm{ex}}(\langle \sigma \rangle)$ & 17.1 & 16.9 & 17.8 
      & 20.1 & 22.5 \\
      $\langle \sigma_I \rangle/\langle \sigma \rangle$ &
      1 & 0.986 & 0.950 & 0.845 & 0.782 \\
      $- \ln P_{\mathrm{ins}}$ & 7.92(1) & 8.098(9) & 8.77(2) & 
      13.68(4) & 26.1(2) \\
      $\ygrow$ & & 32.2(1) & 30.8(2) &29.5(2) & 40.5(1) \\
      $x_V$ & $1.10(2)\cdot10^{-4}$ & $9.55(9)\cdot10^{-5}$ &
      $8.3(2)\cdot10^{-5}$ & $4.6(2)\cdot10^{-5}$ &
      $5(1)\cdot10^{-5}$ \\
      $x_I$ & $2.7(4)\cdot10^{-8}$ & $1.6(2)\cdot10^{-7}$&
      $1.7(3)\cdot10^{-6}$  & $2.4(5)\cdot10^{-3}$ &
      $2.1(2)\cdot10^{-2}$ \\
      \hline
      \hline
    \end{tabular}
  \end{center}
  \caption{Results for the vacancy and interstitial concentration 
    for the polydisperse hard sphere system.
    The interstitial concentration for the monodisperse case was taken from 
    Ref.~\onlinecite{PronkFrenkelPointDefects}.
    All free energies are in units of $k_BT$ and the pressure is in 
    $k_BT/\sigma_0^3$, with the errors in the last digit(s) shown in 
    brackets. 
    Here, $\nu$ is the polydispersity control parameter (see
    Eq.~\ref{pdd:mufunc}), $\eta$ is the packing fraction, $\langle
    \sigma \rangle$ is the mean packing fraction, $s$ is the 
    polydispersity, as defined in Eq.~\ref{pdd:pdispdef}, 
    $\langle \sigma_I \rangle/\langle \sigma \rangle$ is the mean
    interstitial size relative to the mean particle size,
    $P_{\mathrm{ins}}$ is the particle insertion probability (see
    Eq.~\ref{pdd:pins}), $\ygrow$ is the free energy associated with
    growing an interstitial (see Eq.~\ref{pdd:grow}), $x_V$ is the
    vacancy concentration and $x_I$ is the interstitial concentration.
    }
  \label{pdd:restab}
\end{table*}

\section{Discussion}

\begin{figure}
\center
  \includegraphics[width=0.95\linewidth]{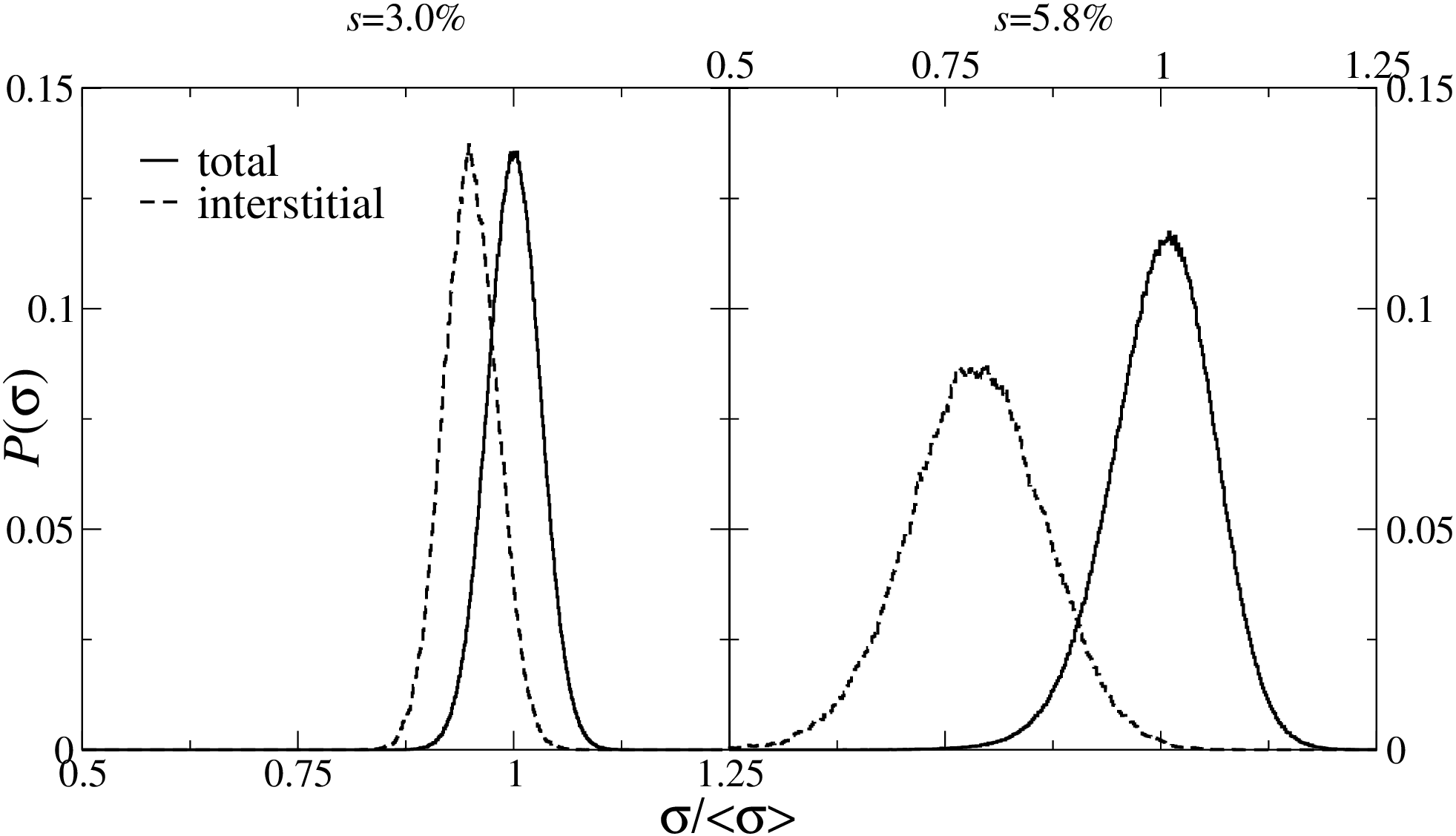}
  \caption{Normalized size distribution for total system and for the 
    interstitials at polydispersities of $3.0\%$ (left, $\nu=0.001$)
    and $5.8\%$ (right, $\nu=0.0056$)}
  \label{pdd:sizedist}
\end{figure}

The simulation results show a dramatic increase in the interstitial
concentration with increasing polydispersity, while the vacancy
concentration remains roughly similar over the full range of
polydispersities. The increase in interstitial concentration can be
attributed to the size of the interstitials: if the particle size
distribution has non-zero width, the interstitials are smaller than
the mean particle size in the crystal, as is shown in
Fig.~\ref{pdd:sizedist}.  

This size difference between interstitials and the surrounding crystal
is not an artifact of the simulation method: although the trial moves
used in semigrand-canonical simulations are unphysical, the resulting
size distribution of interstitials is real. The non-Gaussian particle
size distribution in the crystal should be interpreted as a result of
fractionation\cite{BolhuisPdisp1,FasoloSollichEqPolyHS} of the
coexisting fluid (with the same pressure and polydispersity control
parameter $\nu$).

\begin{figure}
    \center
  \includegraphics[width=0.95\linewidth]{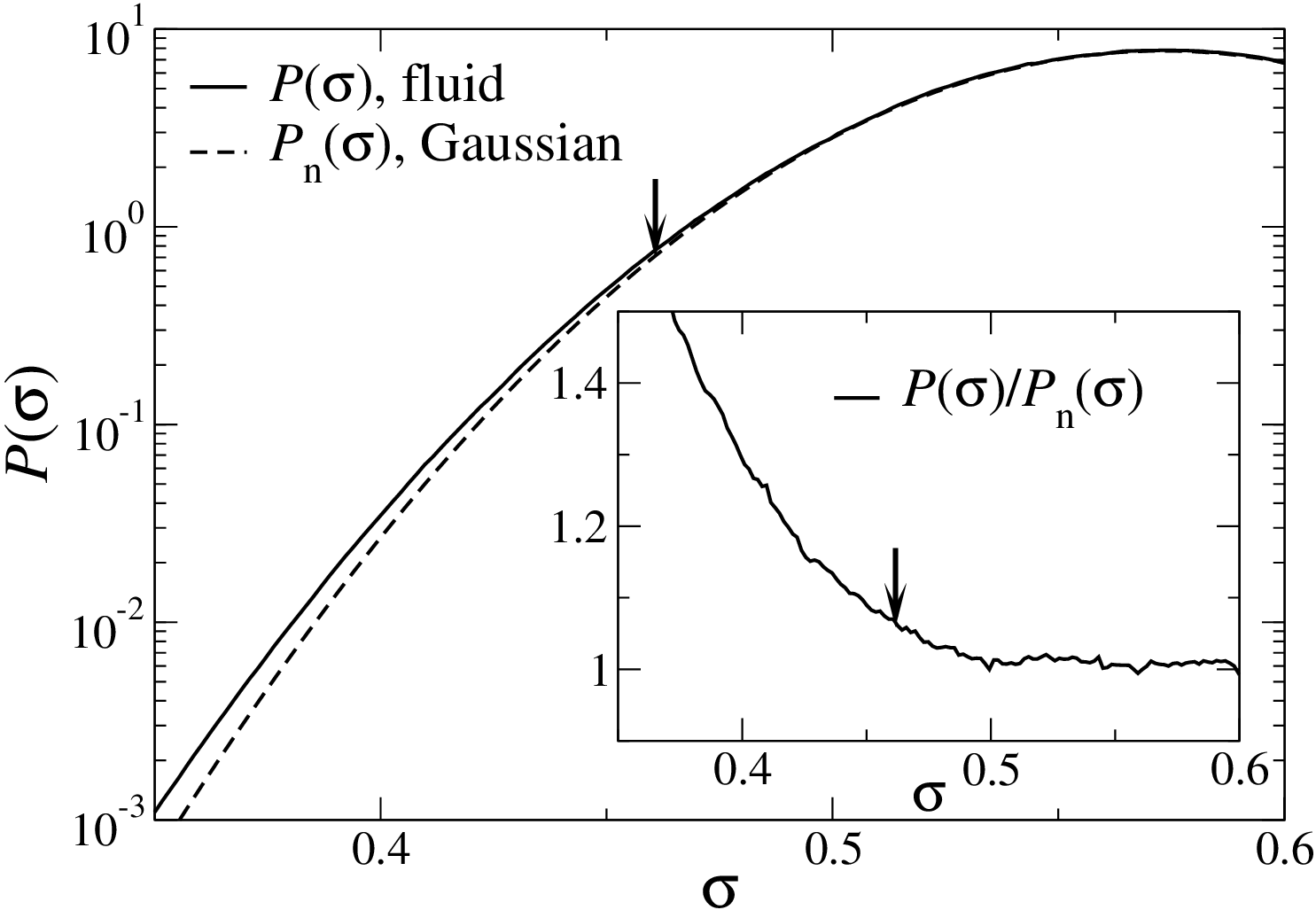}
  \caption{Particle size distribution of the fluid coexisting with the highest
    polydispersity solid (solid line, $\nu=0.0056$, $s=0.058$). The
    dashed line shows a normal distribution with the same first and
    second moment. The inset shows the probability distribution
    relative to the normal distribution. The vertical arrows mark the
    mean particle diameter of interstitials at the current
    polydispersity.}
  \label{pdd:fluidsd}
\end{figure}

The size distribution of the coexisting fluid is shown in
Fig.~\ref{pdd:fluidsd}. For small particle sizes, its value is
slightly higher than the normal distribution, but at the peak of the
interstitial size distribution, the difference in concentration is no
more than 7\%. To a first approximation, the interstitial
concentration in a crystal that has formed from a fluid in which the
particle size distribution is exactly Gaussian, should be lower by the
same amount. 

It must be noted, however, that once the crystalline phase starts
occupying a sizable fraction of the system volume the size
distribution will change and the interstitial concentration will
probably be lower. However, the exact size distribution in the
crystalline phase is difficult to predict; the size distribution of
the fluid itself will change as a result of the growth of the
crystalline phase, and because of the high polydispersity of the
coexisting fluid, the crystalline phase may be composed of several
crystallites, each of which will have its own size distribution.

\begin{figure}
\center
  \includegraphics[width=0.95\linewidth]{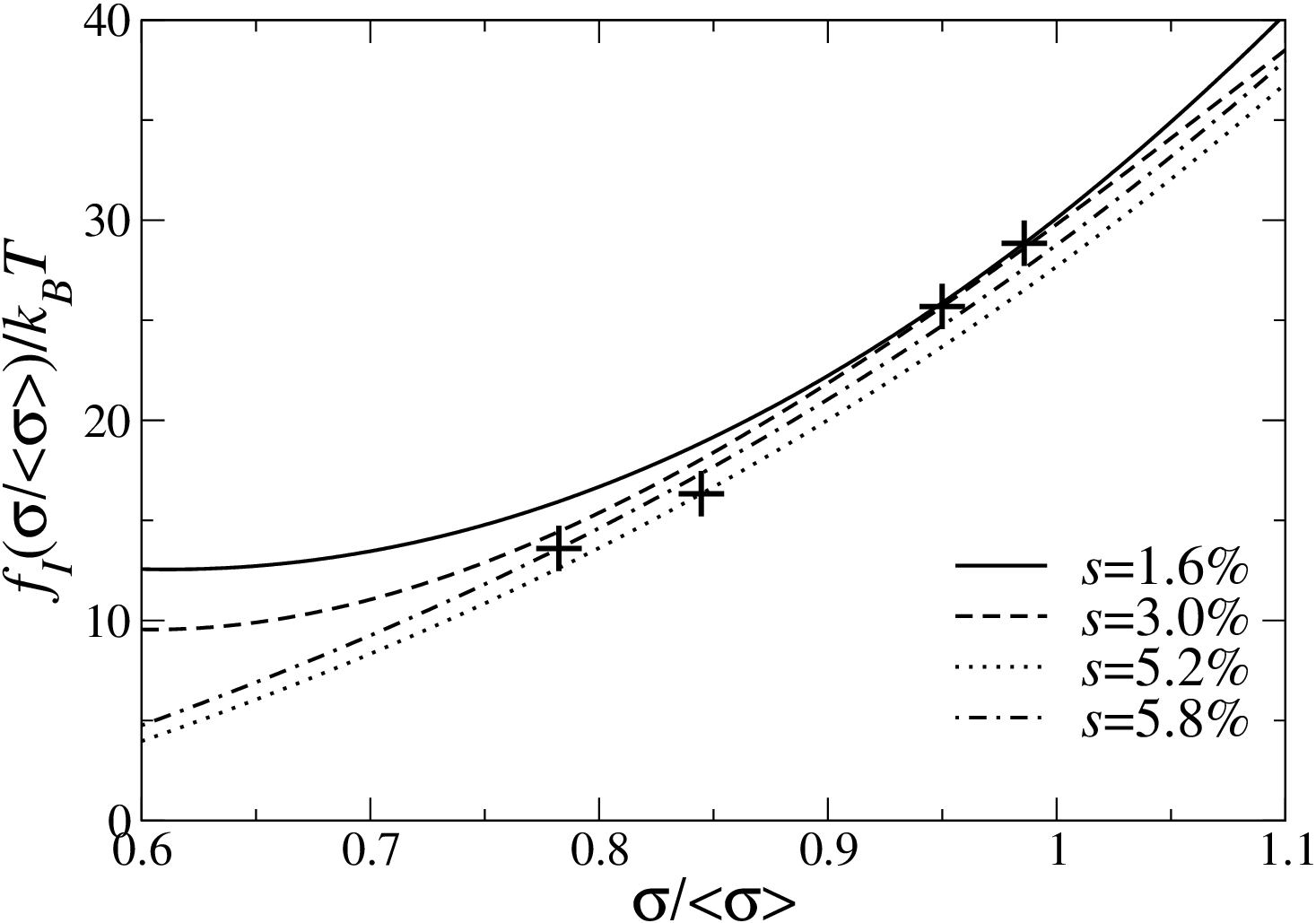}
  \caption{Interstitial free energy $f_I(\sigma)$ for different 
    polydispersities $s$ as a function of renormalized particle size
    $\sigma/\langle \sigma \rangle$. The $\nu$ values for the
    different polydispersities can be found in Table \ref{pdd:restab}.
    The crosses denote the means of the interstitial sizes for the
    corresponding polydispersities.}
  \label{pdd:figraph}
\end{figure}

The influence of the small particles on the interstitial concentration
can be illustrated by looking at the free energy of formation of a
vacancy as a function of size. If we define a partial interstitial
concentration $x_I(\sigma)$, we can, as in Eq.~\ref{pdd:yint}, express
it in terms of the free energy of formation $f_I(\sigma)$ and the
chemical potential:
\begin{equation}
  x_I(\sigma) = \exp \left( - \beta \left[
      f_I(\sigma) - \mu_{\mathrm{ex}}(\sigma)
  \right]
  \right)
 \label{pdd:partconc}
\end{equation}
Assuming that the total interstitial concentration is the integral of
the partial concentrations:
\begin{equation}
 x_I= \int_0^\infty \di \sigma x_I(\sigma) 
 \label{pdd:fullconc}
\end{equation}
we can extract $f_I(\sigma)$, the free energy associated with creating
an interstitial of size $\sigma$, because we know the chemical
potential distribution and the partial interstitial concentration. The
values for $f_I(\sigma)$ at the polydispersities from
Table~\ref{pdd:restab} are shown in Fig~\ref{pdd:figraph}. To be able
to compare values of $f_I(\sigma)$ over a large range of
$\sigma/\langle \sigma \rangle$, the values for $x_I(\sigma)$ in this
figure were obtained by fitting the values from the simulations with
locally skewed Gaussians
\begin{equation}
  x_I(\sigma)
\approx a \exp \left[-b(\sigma-\langle \sigma_I \rangle)^2
    -c(\sigma-\langle \sigma_I \rangle)^3 \right]
\end{equation}
The fits work very well for the values of $\sigma$ which have been
sampled during the simulation, and should yield meaningful results for
the range shown in Fig.~\ref{pdd:figraph}.

The similarity in slopes and actual values of the $f_I(\sigma/\langle
\sigma \rangle)$ 
curves is striking; it means that,
for the full range of polydispersities at which a crystal is stable,
the partial interstitial concentration depends on the chemical
potential distribution and an interstitial free energy which seems to
be only weakly dependent on the polydispersity:
\begin{eqnarray}
  f_I\left(\sigmared\right) & = & 
  - \kappa\left( \frac{1}{2}\sigmared
     - r_0 \right)^2 - f_I^0
  \label{pdd:fitfi}
\end{eqnarray}
with $\kappa=741 k_BT/\sigma_0^2$, $r_0=0.338\sigma_0$ and
$f_I^0=11.3k_BT$ as fitted parameters from the points in
Fig~\ref{pdd:figraph}. Although the form of this equation was taken
from the analytical estimate for the interstitial concentration of
Ref.~\onlinecite{PronkFrenkelPointDefects}, which gives physical
meanings to the values of $\kappa$ and $r_0$ and has reasonable
agreement for $r_0$, we stress that, here, $\kappa$ and $r_0$ are
simply fit parameters.

Because $f_I\left(\sigmaredl\right)$ hardly depends on the width and,
presumably, the shape of the particle size distribution, the small
particle tail of the particle size distribution becomes crucial: those
particles have the lowest $f_I\left(\sigmaredl\right)$ and will form
the most important contribution to the interstitial concentration. For
example, at the near-Gaussian polydispersity of $s=5.2\%$, obtained by
setting $\nu=0.004$, practically all particles with diameter smaller
than $75\%$ of the mean particle radius are interstitials.  This
implies that the polydispersity, as measured by the second moment of
the particle size distribution in the liquid, is \emph{not} a good
predictor for the interstitial concentration in the solid. The
\emph{tail} of the particle size distribution in the liquid is hard to
measure, yet it is all-important for the interstitial concentration.

In the case of vacancies, similar considerations apply in a slightly
different form; the vacancy concentration depends on the chemical
potential and the free energy of removing a particle while keeping its
lattice site. As argued above, they both stay relatively constant at
melting for increasing polydispersities which causes the concentration
of vacancies to remain roughly similar.

To get an estimate for the interstitial concentration of a colloidal
crystal in a suspension, the solely $\sigma$-dependent expression of
Eq.~\ref{pdd:fitfi} must be combined with an estimate for the chemical
potential distribution $\mu_{\mathrm{ex}}(\sigma)$, which, in the more
conventional ensembles of the experimental situation, does not only
depend on the density and the mean particle size, but also on the
subsequent moment of the particle size distribution, the
polydispersity \cite{SollichPdisp,BartlettPolyHSCryst,
  FasoloSollichEqPolyHS}. An estimate for the absolute values of the
chemical potential distribution can be obtained by combining
Eq.~\ref{pdd:mufunc} and the results of table \ref{pdd:restab}.

\bigskip

In summary, we have shown that for polydisperse hard-sphere crystals
along the melting curve, the interstitial concentration increases
dramatically (going up to 2\%) while the vacancy concentration remains
relatively constant. This can be attributed to the fact that, with
increasing polydispersity, there is an increasing probability of
finding a particle small enough to have an appreciable probability of
fitting in a hole of the underlying crystalline lattice.

This finding has practical implication for the preparation of
colloidal crystals from slightly polydisperse solutions. As the
presence of interstitials may affect the optical properties of
colloidal crystals, it is important to control their concentration.
The present calculations show that the interstitial concentration
depends sensitively on the {\em tail} of the size distribution in the
liquid phase. Hence, the polydispersity as such does not provide a
reliable criterion to predict interstitial concentrations. Rather, it
will be necessary to have an accurate representation of the functional
form of the tail of the particle-size distribution (in particular, on
the small-$\sigma$ side).

The work of the FOM institute is part of the research program of the
Foundation for Fundamental Research on Matter (FOM) and was made
possible through financial support by the Dutch Foundation for
Scientific Research (NWO). 

%\bibliography{../main.bib}

\begin{thebibliography}{11}
\expandafter\ifx\csname natexlab\endcsname\relax\def\natexlab#1{#1}\fi
\expandafter\ifx\csname bibnamefont\endcsname\relax
  \def\bibnamefont#1{#1}\fi
\expandafter\ifx\csname bibfnamefont\endcsname\relax
  \def\bibfnamefont#1{#1}\fi
\expandafter\ifx\csname citenamefont\endcsname\relax
  \def\citenamefont#1{#1}\fi
\expandafter\ifx\csname url\endcsname\relax
  \def\url#1{\texttt{#1}}\fi
\expandafter\ifx\csname urlprefix\endcsname\relax\def\urlprefix{URL }\fi
\providecommand{\bibinfo}[2]{#2}
\providecommand{\eprint}[2][]{\url{#2}}

\bibitem[{\citenamefont{van Blaaderen et~al.}(2001)\citenamefont{van Blaaderen,
  Velikov, Hoogenboom, Vossen, Yethiraj, Dullens, van Dillen, and
  Polman}}]{vanBlaaderenPhotonic}
\bibinfo{author}{\bibfnamefont{A.}~\bibnamefont{van Blaaderen}},
  \bibinfo{author}{\bibfnamefont{K.~P.} \bibnamefont{Velikov}},
  \bibinfo{author}{\bibfnamefont{J.~P.} \bibnamefont{Hoogenboom}},
  \bibinfo{author}{\bibfnamefont{D.~L.~J.} \bibnamefont{Vossen}},
  \bibinfo{author}{\bibfnamefont{A.}~\bibnamefont{Yethiraj}},
  \bibinfo{author}{\bibfnamefont{R.~P.~A.} \bibnamefont{Dullens}},
  \bibinfo{author}{\bibfnamefont{T.}~\bibnamefont{van Dillen}},
  \bibnamefont{and} \bibinfo{author}{\bibfnamefont{A.}~\bibnamefont{Polman}},
  in \emph{\bibinfo{booktitle}{Photonic crystals and light organization in the
  21st century}}, edited by \bibinfo{editor}{\bibfnamefont{C.~M.}
  \bibnamefont{Soukoulis}} (\bibinfo{publisher}{Kluwer Academic Publishers},
  \bibinfo{address}{Dordrecht}, \bibinfo{year}{2001}), pp.
  \bibinfo{pages}{239--251}.

\bibitem[{\citenamefont{Kofke and Glandt}(1988)}]{KofkeSemigrand}
\bibinfo{author}{\bibfnamefont{D.~A.} \bibnamefont{Kofke}} \bibnamefont{and}
  \bibinfo{author}{\bibfnamefont{E.~D.} \bibnamefont{Glandt}},
  \bibinfo{journal}{Mol.~Phys} \textbf{\bibinfo{volume}{64}},
  \bibinfo{pages}{1105} (\bibinfo{year}{1988}).

\bibitem[{\citenamefont{Bolhuis and Kofke}(1996)}]{BolhuisPdisp1}
\bibinfo{author}{\bibfnamefont{P.~G.} \bibnamefont{Bolhuis}} \bibnamefont{and}
  \bibinfo{author}{\bibfnamefont{D.~A.} \bibnamefont{Kofke}},
  \bibinfo{journal}{Phys.~Rev.~E} \textbf{\bibinfo{volume}{54}},
  \bibinfo{pages}{634} (\bibinfo{year}{1996}).

\bibitem[{\citenamefont{Polson et~al.}(2000)\citenamefont{Polson, Trizac,
  Pronk, and Frenkel}}]{PolsonEinstein}
\bibinfo{author}{\bibfnamefont{J.~M.} \bibnamefont{Polson}},
  \bibinfo{author}{\bibfnamefont{E.}~\bibnamefont{Trizac}},
  \bibinfo{author}{\bibfnamefont{S.}~\bibnamefont{Pronk}}, \bibnamefont{and}
  \bibinfo{author}{\bibfnamefont{D.}~\bibnamefont{Frenkel}},
  \bibinfo{journal}{J.~Chem.~Phys.} \textbf{\bibinfo{volume}{112}},
  \bibinfo{pages}{5339} (\bibinfo{year}{2000}).

\bibitem[{\citenamefont{Pronk and Frenkel}(2001)}]{PronkFrenkelPointDefects}
\bibinfo{author}{\bibfnamefont{S.}~\bibnamefont{Pronk}} \bibnamefont{and}
  \bibinfo{author}{\bibfnamefont{D.}~\bibnamefont{Frenkel}},
  \bibinfo{journal}{J.~Phys.~Chem.~B} \textbf{\bibinfo{volume}{105}},
  \bibinfo{pages}{6722} (\bibinfo{year}{2001}).

\bibitem[{\citenamefont{Berg and Neuhaus}(1991)}]{BergNeuhausMulticanonical}
\bibinfo{author}{\bibfnamefont{B.~A.} \bibnamefont{Berg}} \bibnamefont{and}
  \bibinfo{author}{\bibfnamefont{T.}~\bibnamefont{Neuhaus}},
  \bibinfo{journal}{Phys.~Rev.~Lett.} \textbf{\bibinfo{volume}{68}},
  \bibinfo{pages}{9} (\bibinfo{year}{1991}).

\bibitem[{\citenamefont{Smith and Bruce}(1996)}]{SmithBruceMulticanonical}
\bibinfo{author}{\bibfnamefont{G.~R.} \bibnamefont{Smith}} \bibnamefont{and}
  \bibinfo{author}{\bibfnamefont{A.~D.} \bibnamefont{Bruce}},
  \bibinfo{journal}{Phys.~Rev.~E} \textbf{\bibinfo{volume}{53}},
  \bibinfo{pages}{6530} (\bibinfo{year}{1996}).

\bibitem[{\citenamefont{Bennett and Alder}(1971)}]{BennettAlderVacancies}
\bibinfo{author}{\bibfnamefont{C.~H.} \bibnamefont{Bennett}} \bibnamefont{and}
  \bibinfo{author}{\bibfnamefont{B.~J.} \bibnamefont{Alder}},
  \bibinfo{journal}{J.~Chem.~Phys.} \textbf{\bibinfo{volume}{54}},
  \bibinfo{pages}{4796} (\bibinfo{year}{1971}).

\bibitem[{\citenamefont{Fasolo and Sollich}(2003)}]{FasoloSollichEqPolyHS}
\bibinfo{author}{\bibfnamefont{M.}~\bibnamefont{Fasolo}} \bibnamefont{and}
  \bibinfo{author}{\bibfnamefont{P.}~\bibnamefont{Sollich}},
  \bibinfo{journal}{Phys.~Rev.~Lett.} \textbf{\bibinfo{volume}{91}},
  \bibinfo{pages}{068301} (\bibinfo{year}{2003}).

\bibitem[{\citenamefont{Sollich}(2002)}]{SollichPdisp}
\bibinfo{author}{\bibfnamefont{P.}~\bibnamefont{Sollich}},
  \bibinfo{journal}{J.~Phys.~Cond.~Matt.} \textbf{\bibinfo{volume}{14}},
  \bibinfo{pages}{1683} (\bibinfo{year}{2002}).

\bibitem[{\citenamefont{Bartlett}(1998)}]{BartlettPolyHSCryst}
\bibinfo{author}{\bibfnamefont{P.}~\bibnamefont{Bartlett}},
  \bibinfo{journal}{J.~Chem.~Phys.} \textbf{\bibinfo{volume}{109}},
  \bibinfo{pages}{10970} (\bibinfo{year}{1998}).

\end{thebibliography}

\end{document}